\documentclass[12pt]{article}
\usepackage{amssymb}
\usepackage{graphicx}
\begin{document}
\newcommand{\beq}{\begin{equation}}
\newcommand{\eeq}{\end{equation}}
\newcommand{\beqa}{\begin{eqnarray}}
\newcommand{\eeqa}{\end{eqnarray}}
\newcommand{\beqar}{\begin{eqnarray*}}
\newcommand{\eeqar}{\end{eqnarray*}}
\newcommand{\al}{\alpha}
\newcommand{\be}{\beta}
\newcommand{\del}{\delta}
\newcommand{\D}{\Delta}
\newcommand{\eps}{\epsilon}
\newcommand{\ga}{\gamma}
\newcommand{\Ga}{\Gamma}
\newcommand{\ka}{\kappa}
\newcommand{\nn}{\nonumber}
\newcommand{\inn}{\!\cdot\!}
\newcommand{\h}{\eta}
\newcommand{\ii}{\iota}
\newcommand{\kk}{\varphi}
\newcommand\F{{}_3F_2}
\newcommand{\la}{\lambda}
\newcommand{\La}{\Lambda}
\newcommand{\na}{\prt}
\newcommand{\Om}{\Omega}
\newcommand{\om}{\omega}
\newcommand{\p}{\Phi}
\newcommand{\sig}{\sigma}
\renewcommand{\t}{\theta}
\newcommand{\z}{\zeta}
\newcommand{\ssc}{\scriptscriptstyle}
\newcommand{\eg}{{\it e.g.,}\ }
\newcommand{\ie}{{\it i.e.,}\ }
\newcommand{\labell}[1]{\label{#1}} 
\newcommand{\reef}[1]{(\ref{#1})}
\newcommand\prt{\partial}
\newcommand\veps{\varepsilon}
\newcommand{\pol}{\varepsilon}
\newcommand\vp{\varphi}
\newcommand\ls{\ell_s}
\newcommand\cF{{\cal F}}
\newcommand\cA{{\cal A}}
\newcommand\cS{{\cal S}}
\newcommand\cT{{\cal T}}
\newcommand\cV{{\cal V}}
\newcommand\cL{{\cal L}}
\newcommand\cM{{\cal M}}
\newcommand\cN{{\cal N}}
\newcommand\cG{{\cal G}}
\newcommand\cK{{\cal K}}
\newcommand\cH{{\cal H}}
\newcommand\cI{{\cal I}}
\newcommand\cJ{{\cal J}}
\newcommand\cl{{\iota}}
\newcommand\cP{{\cal P}}
\newcommand\cQ{{\cal Q}}
\newcommand\cg{{\tilde {{\cal G}}}}
\newcommand\cR{{\cal R}}
\newcommand\cB{{\cal B}}
\newcommand\cO{{\cal O}}
\newcommand\tcO{{\tilde {{\cal O}}}}
\newcommand\bz{\bar{z}}
\newcommand\bb{\bar{b}}
\newcommand\ba{\bar{a}}
\newcommand\bg{\bar{g}}
\newcommand\bc{\bar{c}}
\newcommand\bw{\bar{w}}
\newcommand\bX{\bar{X}}
\newcommand\bK{\bar{K}}
\newcommand\bA{\bar{A}}
\newcommand\bZ{\bar{Z}}
\newcommand\bF{\bar{F}}
\newcommand\bxi{\bar{\xi}}
\newcommand\bphi{\bar{\phi}}
\newcommand\bpsi{\bar{\psi}}
\newcommand\bprt{\bar{\prt}}
\newcommand\bet{\bar{\eta}}
\newcommand\btau{\bar{\tau}}
\newcommand\hF{\hat{F}}
\newcommand\hA{\hat{A}}
\newcommand\hT{\hat{T}}
\newcommand\htau{\hat{\tau}}
\newcommand\hD{\hat{D}}
\newcommand\hf{\hat{f}}
\newcommand\hK{\hat{K}}
\newcommand\hg{\hat{g}}
\newcommand\hp{\hat{\Phi}}
\newcommand\hi{\hat{i}}
\newcommand\ha{\hat{a}}
\newcommand\hb{\hat{b}}
\newcommand\hQ{\hat{Q}}
\newcommand\hP{\hat{\Phi}}
\newcommand\hS{\hat{S}}
\newcommand\hX{\hat{X}}
\newcommand\tL{\tilde{\cal L}}
\newcommand\hL{\hat{\cal L}}
\newcommand\tG{{\tilde G}}
\newcommand\tg{{\tilde g}}
\newcommand\tphi{{\widetilde \Phi}}
\newcommand\tPhi{{\widetilde \Phi}}
\newcommand\te{{\tilde e}}
\newcommand\tk{{\tilde k}}
\newcommand\tf{{\tilde f}}
\newcommand\ta{{\tilde a}}
\newcommand\tb{{\tilde b}}
\newcommand\tc{{\tilde c}}
\newcommand\td{{\tilde d}}
\newcommand\tm{{\tilde m}}
\newcommand\tmu{{\tilde \mu}}
\newcommand\tnu{{\tilde \nu}}
\newcommand\talpha{{\tilde \alpha}}
\newcommand\tbeta{{\tilde \beta}}
\newcommand\trho{{\tilde \rho}}
 \newcommand\tR{{\tilde R}}
\newcommand\teta{{\tilde \eta}}
\newcommand\tF{{\widetilde F}}
\newcommand\tK{{\tilde K}}
\newcommand\tE{{\widetilde E}}
\newcommand\tpsi{{\tilde \psi}}
\newcommand\tX{{\widetilde X}}
\newcommand\tD{{\widetilde D}}
\newcommand\tO{{\widetilde O}}
\newcommand\tS{{\tilde S}}
\newcommand\tB{{\tilde B}}
\newcommand\tA{{\widetilde A}}
\newcommand\tT{{\widetilde T}}
\newcommand\tC{{\widetilde C}}
\newcommand\tV{{\widetilde V}}
\newcommand\thF{{\widetilde {\hat {F}}}}
\newcommand\Tr{{\rm Tr}}
\newcommand\tr{{\rm tr}}
\newcommand\STr{{\rm STr}}
\newcommand\hR{\hat{R}}
\newcommand\M[2]{M^{#1}{}_{#2}}
\newcommand\MZ{\mathbb{Z}}
\newcommand\MR{\mathbb{R}}
\newcommand\bS{\textbf{ S}}
\newcommand\bI{\textbf{ I}}
\newcommand\bJ{\textbf{ J}}

\begin{titlepage}
\begin{center}

\vskip 2 cm
{\LARGE \bf  
Exploring types I and IIA effective actions  \\  \vskip 0.25 cm  through T-duality
 }\\
\vskip 1.25 cm
 Mohammad R. Garousi\footnote{garousi@um.ac.ir}

\vskip 1 cm
{{\it Department of Physics, Faculty of Science, Ferdowsi University of Mashhad\\}{\it P.O. Box 1436, Mashhad, Iran}\\}
\vskip .1 cm
 \end{center}

\begin{abstract}

 It is well-established that compactifying type I string theory on a circle \( S^{(1)} \) transforms the theory under T-duality into type I' theory—the compactification of type IIA string theory on the orbifold \( \tilde{S}^{(1)}/\mathbb{Z}_2 \), where the \( \mathbb{Z}_2 \) action combines worldsheet parity with spacetime reflection along the dual circle \( \tilde{S}^{(1)} \). We propose that, upon compactification, the untwisted (twisted) sector of the type I effective action should map under the Buscher rules to the untwisted (twisted) sector of the type I' effective action. This T-duality constraint offers significant insight into the determination of bosonic couplings in the effective action of type IIA theories, specifically those that remain after orbifold reduction, as well as in the untwisted sector of the type I effective action. However, its scope is limited and insufficient to fully determine the couplings within the twisted sectors of type I and type I' theories.  Within this framework, we demonstrate that the leading 2-derivative couplings in untwisted sector of type I and the 2-derivative couplings in type IIA  theory are uniquely determined, except for the Chern-Simons term in type IIA, which is absent in the orbifold reduction.


\end{abstract}

\end{titlepage}


String theory is a promising candidate for quantum gravity, as it encompasses both massless fields and an infinite tower of massive fields. At low energies, it is described by an effective action in  10-dimensional spacetime that includes only the massless fields and their derivatives, classified in terms of the sole dimensionful parameter in string theory, $\alpha'$. The couplings at each order of $\alpha'$ include classical couplings \cite{Gross:1986iv,Gross:1986mw,Grisaru:1986kw,Grisaru:1986vi}, loop or genus corrections \cite{Sakai:1986bi,Ellis:1987dc,Abe:1988cq,Abe:1987ud}, as well as non-perturbative corrections \cite{Green:1997tv,Green:1997di,Kiritsis:1997em,Pioline:1997pu}. This effective action is expected to serve as a viable framework for quantum gravity in  10-dimensional spacetime.

The  effective action in string theory can be constructed by imposing various global and local symmetries inherent to the theory. While the leading-order effective action can be determined by requiring local supersymmetry, extending this approach to higher-order derivative terms is significantly more challenging. This complexity arises because one must simultaneously account for both bosonic and fermionic fields in a consistent manner (see \cite{Ozkan:2024euj} for a comprehensive review). Among the notable global symmetries in spacetime is the $O(d,d,\MR)$ symmetry, which emerges when the 10-dimensional classical effective action is dimensionally reduced on a torus  $T^{(d)}$ \cite{Sen:1991zi,Hohm:2014sxa}. 
 This symmetry is based on the implicit assumption that the effective action is background-independent. This means the structure of the  effective action remains consistent and unchanged, irrespective of whether certain dimensions are flat or compactified on circles.  At the classical level, effective actions are indeed background-independent \cite{Garousi:2022ovo}; however, at the quantum level, this independence is lost. For instance, S-matrix calculations at the torus level are sensitive to the presence or absence of a circular dimension in spacetime \cite{Green:1982sw}. The introduction of a circular dimension leads to two significant effects. First, the Kaluza-Klein momenta in loops become discrete, replacing the continuous integral over internal momenta with a summation. Second, winding momenta emerge, a feature absent in spacetimes without a circular dimension. 
These effects introduce a dependence of S-matrix elements on the background topology, rendering quantum corrections to the effective action no longer background-independent. Furthermore, under T-duality, the Kaluza-Klein momenta and winding momenta are interchanged. As a result, while the classical effective action exhibits the full \( O(d,d,\mathbb{R}) \) symmetry, quantum corrections—such as higher-genus contributions—are confined to the discrete subgroup \( O(d,d,\mathbb{Z}) \). Naturally, the classical effective action also respects \( O(d,d,\mathbb{Z}) \), since this subgroup is inherently part of the continuous symmetry.

It has been shown that imposing $O(1,1,\mathbb{Z})$ symmetry on the circular reduction of the classical effective action allows for the derivation of all couplings in the effective action of bosonic string theory, as well as all bosonic couplings in the effective action of heterotic string theory at orders $\alpha'$ and $\alpha'^2$ \cite{Kaloper:1997ux,Garousi:2019wgz,Garousi:2019mca,Garousi:2024avb,Garousi:2024imy}. Additionally, all classical NS-NS couplings in type II superstring theories at order $\alpha'^3$ have been identified in \cite{Garousi:2020gio}.
Moreover, the R-R couplings in these theories at the leading order of $\alpha'$ have been shown in \cite{Hassan:1999bv,Garousi:2019jbq} to align with $O(1,1,\mathbb{Z})$ transformations, which relate the type IIA classical couplings to their corresponding type IIB couplings. All of the aforementioned classical couplings are fixed up to overall multiplicative factors in each case. Once these factors are determined for Minkowski spacetime, they remain valid for any other spacetime, owing to the background independence of the classical effective action.
On the other hand, since quantum corrections to the effective action are background-dependent but still invariant under $O(1,1,\mathbb{Z})$ transformations, the above results extend to quantum corrections as well. However, the overall factors for quantum corrections must be calculated using loop-level S-matrix computations for a specific background with one Killing circle. These results are then applicable exclusively to such spacetimes \cite{Garousi:2025xdz}.

It is well-established that compactifying type I string theory on the circle $S^{(1)}$, under T-duality, transforms it into type I' string theory. The latter corresponds to the compactification of type IIA string theory on the orbifold $\tilde{S}^{(1)}/\mathbb{Z}_2$, where $\mathbb{Z}_2$ combines world-sheet parity with spacetime parity along the dual circle $\tilde{S}^{(1)}$ \cite{Schwarz:1999xj}. Both type I and type I' theories consist of untwisted and twisted sectors.
The untwisted sector of the type I' effective action shares the same world-sheet topologies as the type IIA effective action, whereas the twisted sector encompasses all other world-sheet topologies. Under T-duality, the untwisted (twisted) sector of the effective action in type I string theory maps to the untwisted (twisted) sector of the effective action in type I' string theory.
This paper focuses exclusively on the untwisted sector. Specifically, we aim to demonstrate that by applying $O(1,1,\mathbb{Z})$ transformations or Buscher rules to the circular reduction of the untwisted sector of the type I effective action, one can derive the effective action of type IIA string theory. This approach provides strong constraints for determining higher-derivative couplings in the untwisted sector of the type I effective action and identifying couplings in the type IIA classical effective action that remain non-vanishing under orbifold reduction. 

On the other hand, the twisted sector of the type I effective action is 10-dimensional, whereas, for a zero Wilson line in type I theory, the twisted sector of type I' theory is restricted to a single 9-dimensional fixed point of the orbifold. Consequently, as we will demonstrate, the T-duality constraint on the twisted sector is insufficient to fully determine the unknown coupling constants in the twisted sector of the 10-dimensional type I theory, as well as the independent coupling constants in the twisted sector of type I' theory. Notably, since the twisted sector of the type I' effective action is inherently 9-dimensional, it includes numerous coupling constants.  Therefore, the T-duality constraint alone cannot fully resolve these couplings.
This contrasts with the untwisted sector, where the independent couplings in both theories are 10-dimensional, resulting in significantly fewer unknown coupling constants that emerge in 9-dimensional spacetime after compactification. To determine the twisted sector using T-duality, one would need to independently identify the couplings in the twisted sector of type I theory through alternative methods—an investigation that falls outside the scope of this paper.

In the type I effective action, specific couplings arise from the replacement of deformed Ramond-Ramond (R-R) field strength, as required by the Green-Schwarz mechanism \cite{Green:1984sg}, within the untwisted sector of the theory. These couplings are notably absent in the type IIA effective action due to its fundamentally non-chiral nature. Our analysis indicates that all such couplings in type I theory reside exclusively in the twisted sector of the effective action. 
A subtlety arises in the process of dimensional reduction: although the 10-dimensional type IIA theory is non-chiral, its orbifold reduction results in a chiral theory \cite{Schwarz:1999xj}. Crucially, all terms associated with potential chiral anomalies are confined to the twisted sector, while the untwisted sector remains anomaly-free.

The massless bosonic fields of type I theory are divided into two sectors: the twisted sector, which includes massless open string fields $A_\alpha^a$ (not the focus of this paper), and the untwisted sector, which consists of closed string fields. These closed string fields can be derived from the massless fields of type IIB string theory by gauging the  world-sheet parity symmetry of type IIB theory. The massless fields in type IIB theory include $G_{\alpha\beta}$, $B_{\alpha\beta}$, and $\Phi$ in the NS-NS sector, and $C^{(0)}$, $C^{(2)}$, and $C^{(4)}$ in the R-R sector, with the $C^{(4)}$-form being self-dual. The fields that survive the orientifold projection  are:
\beqa
G_{\alpha\beta}, \Phi, C^{(2)},\labell{typeI}
\eeqa
which are the massless closed string fields of 10-dimensinal type I string theory. 

To study the T-duality of type I theory, one should consider one of the spatial coordinates of spacetime to be a circle. If the circle has the coordinate $y$, then the fields mentioned above have the following 9-dimensional components:
\beqa
G_{\mu\nu}, G_{\mu y}, G_{yy}, \Phi, C_{\mu\nu}, C_{\mu y}\,.\labell{I}
\eeqa
On the other hand, the 10-dimensional type IIA theory has the following massless  bosonic fields:
\beqa
G_{\alpha\beta}, B_{\alpha\beta}, \Phi, C^{(1)}, C^{(3)}\,.\labell{typeIIA}
\eeqa
Under the orbifold reduction, the surviving components of these massless fields in the 9-dimensional type I' are:
\beqa
G_{\mu\nu}, G_{yy}, B_{\mu y}, \Phi, C_{\mu}, C_{\mu \nu y}\,.\labell{I'}
\eeqa
Note that the number of massless fields in \reef{I} is the same as the number of massless fields in \reef{I'}. The transformation of the circular reduction of type I fields \reef{I} under the Buscher rules results in the following fields in type I' \cite{Buscher:1987sk,Rocek:1991ps,Meessen:1998qm}:
\beqa
&&G_{\mu\nu}\rightarrow G_{\mu\nu}+\frac{B_{\mu y}B_{\nu y}}{G_{yy}}\,\,\,,\,\,\,G_{\mu y}\rightarrow\frac{B_{\mu y}}{G_{yy}}\,\,\,,\,\,\,
G_{yy}\rightarrow\frac{1}{G_{yy}}\,\,\,,\,\,\,e^{2\Phi}\rightarrow\frac{e^{2\Phi}}{G_{yy}}\nn\\
&& C_{\mu y}\rightarrow C_{\mu}\,\,\,,\,\,\,C_{\mu\nu}\rightarrow C_{\mu\nu y}+C_\mu B_{\nu y}-C_{\nu}B_{\mu y}\,,\labell{treeII}
\eeqa
where we have used in the Buscher rules  \cite{Buscher:1987sk,Rocek:1991ps,Meessen:1998qm} the fact that type I' theory has no $G_{\mu y}$.

To derive the classical effective actions of type IIA and the untwisted sector of type I through T-duality, we formulate the most general covariant and gauge-invariant effective action at tree level and at the two-derivative order for these actions. The construction proceeds as follows:
\beqa
\bS_{I}^{(0)}
&\!\!\!\!=\!\!\!\!& -\frac{2}{\kappa^2}\int d^{10}x \sqrt{-G} \Big[e^{-2\Phi}\Big(a_1R + a_2\nabla_\alpha\Phi\nabla^\alpha\Phi\Big) + a_3|F^{(3)}|^2\Big],\labell{S0b}\\
\bS_{IIA}^{(0)}&\!\!\!\!=\!\!\!\!& -\frac{2}{\kappa^2}\int d^{10}x \sqrt{-G} \Big[e^{-2\Phi}\Big(b_1R + b_2\nabla_\alpha\Phi\nabla^\alpha\Phi+b_3 H_{\alpha\beta\gamma} H^{\alpha\beta\gamma}\Big) + b_4|F^{(2)}|^2+ b_5|F^{(4)}|^2\Big],\nn
\eeqa
where $|F^{(n)}|^2 = \frac{1}{n!}F_{\alpha_1\cdots\alpha_n} F^{\alpha_1\cdots\alpha_n}$. Moreover, $H = dB$, $F^{(2)} = dC^{(1)}$, $F^{(3)} = dC^{(2)}$, and $F^{(4)} = dC^{(3)} + H \wedge C^{(1)}$. By rescaling the R-R potential as $C^{(n)} \rightarrow e^{-\Phi} C^{(n)}$, one finds the overall dilaton factor $e^{-2\Phi}$, which corresponds to the sphere-level effective action. However, the above form is a standard representation of the R-R couplings in the literature, and the Buscher rules \reef{treeII} are expressed in terms of the standard R-R fields. The parameters $a_1, a_2, a_3, b_1, b_2, b_3, b_4, b_5$ are coupling constants that we will determine by imposing the constraint that the above classical actions should be covariant under Buscher transformations \reef{treeII}. Specifically, this means $S_{I}^{(0)} \rightarrow S_{I'}^{(0)}$, where $S_I^{(0)}$ is the circular reduction of $\bS_I^{(0)}$ and $S_{I'}^{(0)}$ is the orbifold reduction of $\bS_{IIA}^{(0)}$. There is also a Chern-Simons coupling $\int B \wedge dC^{(3)} \wedge dC^{(3)}$ in the type IIA effective action, which does not survive under the orbifold reduction. Hence, its coupling constant cannot be fixed by the constraint $S_{I}^{(0)} \rightarrow S_{I'}^{(0)}$.

To simplify the Buscher rules \reef{treeII}, we consider the following circular reduction for the type I fields \cite{Maharana:1992my,Garousi:2019jbq}:
\beqa
G_{\alpha\beta}=\left(\matrix{\bg_{\mu\nu}+e^{\varphi}g_{\mu }g_{\nu }& e^{\varphi}g_{\mu }&\cr e^{\varphi}g_{\nu }&e^{\varphi}&}\!\!\!\!\!\right),\, C_{\alpha\beta}=\left(\matrix{\bc_{\mu\nu}+\bc_\mu g_{\nu }-\bc_{\nu}g_\mu& \bc_{\mu }&\cr -\bc_{\nu }&0&}\!\!\!\!\!\right),\,\Phi=\bar{\phi}+\varphi/4\,,\labell{reduc}
\eeqa
and the following orbifold reduction for the type IIA fields:
\beqa
G_{\alpha\beta}=\left(\matrix{\bg_{\mu\nu}&0&\cr0&e^{\varphi}&}\!\!\!\!\!\right),\, B_{\alpha\beta}=\left(\matrix{0& b_{\mu }&\cr -b_{\nu }&0&}\!\!\!\!\!\right),\,\Phi=\bar{\phi}+\varphi/4\,,C_{\alpha}=\left(\matrix{\bc_\mu& \cr 0&}\!\!\!\!\!\right)\,,C_{\alpha\beta\gamma}=\left(\matrix{0& \cr \bc_{\mu\nu}&}\!\!\!\!\!\right).\labell{oreduc}
\eeqa
Then the transformations \reef{treeII} simplify to the following linear transformations:
\beqa
g_\mu\rightarrow b_\mu, \,\,\vp\rightarrow -\vp, \,\,\bphi\rightarrow \bphi,\,\,\bg_{\mu\nu}\rightarrow \bg_{\mu\nu},\,\,\bc_\mu\rightarrow \bc_\mu,\,\,\bc_{\mu\nu}\rightarrow \bc_{\mu\nu}\,.\labell{trans}
\eeqa
Imposing the above transformations on the 9-dimensional type I couplings should yield the corresponding type I' couplings. This serves as a constraint to determine the coupling constants in the leading-order actions \reef{S0b}. While these transformations are expected to receive higher-order corrections—essential for establishing the higher-derivative effective action—this paper does not focus on such corrections. It is worth emphasizing the following transformation:
\beqa
b_\mu\rightarrow g_\mu, \,\,\vp\rightarrow -\vp, \,\,\bphi\rightarrow \bphi,\,\,\bg_{\mu\nu}\rightarrow \bg_{\mu\nu},\,\,\bc_\mu\rightarrow \bc_\mu,\,\,\bc_{\mu\nu}\rightarrow \bc_{\mu\nu}\,,\labell{trans1}
\eeqa
 converts the 9-dimensional type I' couplings to 9-dimensional type I couplings, \ie $S_{I'}^{(0)}\rightarrow S_I^{(0)}$. Hence, the transformation generates the $\MZ_2$-group.

Using the reduction \reef{reduc}, one finds that the circular reduction of the leading-order type I effective action \reef{S0b} in flat base space becomes \cite{Garousi:2019wgz,Garousi:2019jbq}
\beqa
   S_I^{(0)}&=& -\frac{2}{\kappa^2}\int d^{9}x \left(e^{-2\bphi} \,  \Big[(-\frac{1}{2}a_1+\frac{1}{16}a_2)\prt_\mu\vp\prt^\mu\vp+a_2\prt_\mu\bphi\prt^\mu\bphi -a_1 \prt^\mu\prt_\mu\vp+\frac{1}{2}a_2\prt^\mu\bphi\prt_\mu\vp \right.\nn\\
  &&\left.\qquad\qquad\qquad\qquad-\frac{1}{4}a_1e^{\vp}V^2 \Big]+a_3e^{-\vp/2}|\bF^{(2)}|^2+a_3e^{\vp/2}|\bF^{(3)}-\bc^{(1)}\wedge V|^2\right),\labell{RR}
 \eeqa
where $V_{\mu\nu} = \prt_\mu g_\nu - \prt_\nu g_\mu$ represents the field strength of the $U(1)$ vector $g_\mu$, and $\bF^{(n)} = d\bc^{(n-1)}$. Furthermore, by applying the orbifold reduction \reef{oreduc}, the orbifold reduction of the leading order type IIA effective action \reef{S0b} is obtained as \cite{Garousi:2019wgz,Garousi:2019jbq}
\beqa
   S_{I'}^{(0)}&=& -\frac{2}{\kappa^2}\int d^{9}x \left(e^{-2\bphi} \,  \Big[(-\frac{1}{2}b_1+\frac{1}{16}b_2)\prt_\mu\vp\prt^\mu\vp+b_2\prt_\mu\bphi\prt^\mu\bphi -b_1 \prt^\mu\prt_\mu\vp+\frac{1}{2}b_2\prt^\mu\bphi\prt_\mu\vp \right.\nn\\
  &&\left.\qquad\qquad\qquad\qquad+3b_3e^{-\vp}W^2 \Big]+b_4e^{\vp/2}|\bF^{(2)}|^2+b_5e^{-\vp/2}|\bF^{(3)}-\bc^{(1)}\wedge W|^2\right),\labell{RR1}
 \eeqa
where $W_{\mu\nu} = \prt_\mu b_\nu - \prt_\nu b_\mu$ is the field strength of the $U(1)$ vector $b_\mu$. The T-duality constraint is that, up to some total derivative terms in the base space, $S_I^{(0)} \rightarrow S_{I'}^{(0)}$ under the transformation \reef{trans} or $S_{I'}^{(0)} \rightarrow S_I^{(0)}$ under the transformation \reef{trans1}.

Up to a total derivative term, one finds that under the transformations \reef{trans}, the 9-dimensional action \reef{RR} transforms to the 9-dimensional action \reef{RR1}, provided that the following relationships between the coupling constants hold:
\beqa
b_1=a_1,\,b_2=4a_1,\, b_3=-\frac{1}{12}a_1,\, b_4=a_3,\, b_5=a_3,\, a_2=4a_1\,.
\eeqa
Moreover, the constants $a_1$ and $a_3$ can also be absorbed by normalizing the metric $G_{\alpha\beta}$ and the R-R potential $C^{(2)}$ to have standard kinetic terms, \ie $a_1=1$ and $a_3=-1/2$. It is important to emphasize that if the 9-dimensional couplings in \reef{RR1} were treated as independent, \reef{RR1} would contain seven parameters. In this scenario, T-duality would not be able to constrain all the unknown parameters. Therefore, it is essential that the independent couplings in both theories are defined in 10 dimensions, as illustrated in \reef{S0b}, allowing T-duality to impose constraints on their reductions to 9 dimensions. This demonstrates why the T-duality constraint alone, in the twisted sector, cannot uniquely determine the couplings in the twisted sectors of type I and type I' theories. This limitation arises from the fact that the couplings in the twisted sector of type I' are inherently 9-dimensional.

Hence, the type I and type IIA effective actions in \reef{S0b} become
\beqa
\bS_{I}^{(0)}
&\!\!\!\!=\!\!\!\!& -\frac{2}{\kappa^2}\int d^{10}x \sqrt{-G} \Big[e^{-2\Phi}\Big(R + 4\nabla_\alpha\Phi\nabla^\alpha\Phi\Big) -\frac{1}{2}|F^{(3)}|^2\Big],\labell{S0b1}\\
\bS_{IIA}^{(0)}&\!\!\!\!=\!\!\!\!& -\frac{2}{\kappa^2}\int d^{10}x \sqrt{-G} \Big[e^{-2\Phi}\Big(R + 4\nabla_\alpha\Phi\nabla^\alpha\Phi-\frac{1}{12} |H|^2\Big) -\frac{1}{2}|F^{(2)}|^2- \frac{1}{2}|F^{(4)}|^2\Big]\,.\nn
\eeqa
Up to the Chern-Simons term  $\int B \wedge dC^{(3)} \wedge dC^{(3)}$, the action $\bS_{IIA}^{(0)}$ represents the standard leading-order 10-dimensional effective action of non-chiral type IIA string theory. However, $\bS_{I}^{(0)}$ does not yet constitute the  standard 10-dimensional effective action of the chiral type I string theory.

In type I theory, there is a chiral anomaly which can be canceled for the gauge group $SO(32)$ by adding appropriate couplings at the 8-derivative order and by deforming the R-R gauge transformation to include non-standard $A_\alpha{}^a$-gauge transformations and local Lorentz transformations \cite{Green:1984sg}. In this paper, we specifically consider the case of zero gauge field. Under the non-standard local Lorentz transformation for the $C^{(2)}$-field, the $C^{(2)}$-field strength in \reef{S0b1} needs to be replaced by a new field strength that is invariant under these non-standard local Lorentz transformations \cite{Green:1984sg}, \ie
\beqa
F^{(3)}_{\alpha\beta\gamma} &\rightarrow F^{(3)}_{\alpha\beta\gamma} + \frac{3}{2}\alpha' \Omega_{\alpha\beta\gamma},\labell{replace}
\eeqa
where the Chern-Simons three-form $\Omega$ is constructed from the frame $e_{\alpha}{}^{\alpha_1}$, which is related to the spacetime metric as $e_{\alpha}{}^{\alpha_1} e_\beta{}^{\beta_1} \eta_{\alpha_1\beta_1} = G_{\alpha\beta}$. Our index convention is that $\alpha, \beta$ are the indices of the curved spacetime, and $\alpha_1, \beta_1$ are the indices of the flat tangent spaces.

The replacement of \reef{replace} into \reef{S0b1} yields the following terms at orders $\alpha'$ and $\alpha'^2$:
\beqa
\bS^{(1)}_I&=&-\frac{2\alpha'}{\kappa^2}\int d^{10} x \sqrt{-G} \left(\frac{1}{4}F^{(3)}_{\alpha\beta\gamma}\Omega^{\alpha\beta\gamma}\right),\nn\\
\bS^{(2)}_{I}&=&-\frac{2\alpha'^2}{\kappa^2}\int d^{10} x \sqrt{-G} \left(\frac{3}{16}\Omega_{\alpha\beta\gamma}\Omega^{\alpha\beta\gamma}\right)\,.
\labell{CS}
\eeqa
There are no such terms in the 10-dimensional type IIA effective action. However, this does not imply any inconsistency with T-duality. Specifically, if the R-R potential is rescaled such that the overall dilaton factor of the couplings in \reef{S0b1} becomes $e^{-2\Phi}$, this rescaling results in the dilaton factor of $\bS^{(1)}_I$ being $e^{-\Phi}$ and that of $\bS^{(2)}_I$ being 1. It is important to note that the frame $e_{\alpha}{}^{\alpha_1}$ cannot be rescaled in a similar way, as doing so would disrupt the dilaton factor $e^{-2\Phi}$ in the curvature term of \reef{S0b1}.
Consequently, the coupling $\bS^{(1)}_I$ arises from disk and projective plane orders, while the coupling $\bS^{(2)}_I$ originates from cylinder, Möbius, and Klein bottle orders. Furthermore, the torus world-sheet cannot generate non-zero couplings at order $\alpha'^2$ \cite{Ellis:1987dc}. Such world-sheets are absent in type IIA theory. As a result, all these couplings are associated with the twisted sector of type I theory.
This twisted sector is expected to include numerous additional four-derivative and six-derivative couplings. All these couplings must transform under T-duality into the twisted sector of type I' theory. However, as previously mentioned, T-duality alone cannot uniquely determine these couplings, placing them outside the scope of our current investigation.

In the heterotic theory, which lacks a twisted sector, the Green-Schwarz mechanism requires the following replacement within the effective action:
\beqa
H_{\alpha\beta\gamma} &\rightarrow H_{\alpha\beta\gamma} + \frac{3}{2}\alpha' \Omega_{\alpha\beta\gamma}. \labell{Hreplace}
\eeqa
This gives rise to couplings similar to those in \reef{CS}, but with the inclusion of the dilaton factor $e^{-2\Phi}$. These couplings remain consistent with T-duality after incorporating additional 10-dimensional  couplings at orders $\alpha'$ and $\alpha'^2$ \cite{Garousi:2023kxw}. Importantly, all these couplings are uniquely determined by the T-duality constraint. 

The next classical higher derivative corrections to the actions \reef{S0b1} are at the 8-derivative order. To impose T-duality and identify such couplings, one should first determine $\bS_I^{(3)}$, which includes all independent couplings in type I theory involving the fields \reef{typeI}, and $\bS_{IIA}^{(3)}$, which includes all independent couplings in type IIA theory involving the fields \reef{typeIIA}. Then, the circular reduction of type I couplings should be used to find $S_I^{(3)}$, and the orbifold reduction of type IIA couplings should be used to find $S_{I'}^{(3)}$. The T-duality constraint requires that $S_I^{(3)}$ transforms into $S_{I'}^{(3)}$ under the higher derivative corrections of the transformation \reef{trans}.

However, these calculations are quite lengthy due to the numerous independent couplings in types I and  IIA theories. A convenient approach to manage this large number of couplings is to categorize them based on the number of R-R field strengths. First, identify the independent terms with zero R-R field strengths and use T-duality to determine their coupling constants. Next, identify the independent couplings with two R-R field strengths and apply T-duality to determine their coupling constants. This process should continue up to eight R-R field strengths. Similar calculations for determining massless open string couplings on D-brane world-volume actions, based on the number of Maxwell field strengths, have been conducted in \cite{Karimi:2018vaf}. The details of these calculations will be addressed in future work.


\end{document}